\pdfoutput=1
\documentclass[11pt,a4paper]{article}
\usepackage{amsmath,amssymb,graphicx,color,subfigure,enumitem,dsfont}
\usepackage[colorlinks,linkcolor=purple,citecolor=teal]{hyperref}
\usepackage[compress,square,numbers]{natbib}
\usepackage{tikz}
\usetikzlibrary{decorations.pathreplacing}

\newcommand{\beq}{\begin{eqnarray}}
\newcommand{\eeq}{\end{eqnarray}}
\newcommand{\non}{\nonumber\\}

\DeclareMathOperator{\SU}{SU}

\DeclareMathOperator{\SO}{SO}

\newcommand{\p}{\partial}
\renewcommand{\i}{\mathrm{i}}
\renewcommand{\d}{\mathop{}\!\mathrm{d}}

\DeclareMathOperator{\tr}{tr}

\newcommand{\calE}{\mathcal{E}}

\newcommand{\calO}{\mathcal{O}}
\newcommand{\calQ}{\mathcal{Q}}

\newcommand{\bx}{\mathbf{x}}

\newcommand{\du}{\delta u}
\newcommand{\dub}{\delta\bar{u}}

\newcommand{\zb}{\bar{z}}

\newcommand{\bA}{\mathbf{A}}

\newcommand{\bn}{\mathbf{n}}

\newcommand{\ub}{\bar{u}}
\newcommand{\vb}{\bar{v}}

\newcommand{\figsfolder}{}

\makeatletter
\@addtoreset{equation}{section}

\renewcommand{\thefootnote}{\fnsymbol{footnote}}
\makeatother

\makeatletter
\newcommand{\thetablename}{Table}
\def\fnum@table{\thetablename\ \thetable}
\makeatother

\begin{document}
\pagenumbering{Roman}
\thispagestyle{empty}
\vspace{3mm}

\begin{center}
{\Large \bf Creation of domain-wall Skyrmions in chiral magnets} 
\\[15mm]
{Sven Bjarke~{\sc Gudnason}}$^{1}$,
{Yuki~{\sc Amari}}$^{2,3}$,
{Muneto~{\sc Nitta}}$^{3,2,4}$
\vskip 6 mm

\bigskip\bigskip
{\it
${}^1$Institute of Contemporary Mathematics, School of
  Mathematics and Statistics,\\ Henan University, Kaifeng, Henan 475004,
  P.~R.~China\\
${}^2$Department of Physics, and Research and Education Center
  for Natural Sciences,\\ Keio University, Hiyoshi 4-1-1, Yokohama,
  Kanagawa 223-8521, Japan\\
${}^3$Department of Physics, Keio University, 4-1-1 Hiyoshi,
  Yokohama, Kanagawa 223-8521, Japan
${}^4$International Institute for Sustainability with Knotted
  Chiral Meta Matter(WPI-SKCM2), Hiroshima University, 1-3-2 Kagamiyama,
  Higashi-Hiroshima, Hiroshima 739-8511, Japan
}

\bigskip

\bigskip

{\bf Abstract}\\[5mm]
{\parbox{13cm}{\hspace{5mm}
We study the capture of a magnetic Skyrmion into a domain wall (DW)
structure in chiral magnets and find that in the respective
ground states of the DW and the Skyrmion, they repel each other.
This means that an isolated magnetic Skyrmion cannot \emph{a priori} enter the DW and become a DW-Skyrmion.
However, rotating the DW's phase away from the stable phase may cause the successful capture and hence
creation of a stable bound state of a magnetic Skyrmion and a DW: the DW-Skyrmion.
At certain distances, the DW may also destroy the magnetic Skyrmion by inducing shrinkage of the latter.
This happens as the isolated magnetic Skyrmion has negative DMI energy, whereas the DW-Skyrmion has positive DMI energy; in a finite range from the DW, the magnetic Skyrmion can be pushed to have vanishing DMI energy, for which it collapses to a point.
\emph{En passant}, we find the possibility of pair creation of a Skyrmion anti-DW-Skyrmion pair and a creation of more than one DW-Skyrmions by a Kibble-like mechanism.
}}
\end{center}
\newpage
\pagenumbering{arabic}
\setcounter{page}{1}
\setcounter{footnote}{0}
\renewcommand{\thefootnote}{\arabic{footnote}}
%%%%%%%%%%%%%%%%%%%%%%%%%%%%%%%%

%%%%%%%%%%%%%%%%%%%%%%%%%%%%%%%%%%%%%%%%%%%%%%%%%%%%%%%%%%%%%%%%

\tableofcontents

\section{Introduction}

Skyrmions are particle-like topological solitons supported by the third homotopy group 
$\pi_3 (S^3) \simeq {\mathbb Z}$ 
in the Skyrme model 
\cite{Skyrme:1961vq,Skyrme:1962vh,Manton:2022fcb} 
in which baryons are described as Skyrmions. 
The Skyrme model is 
justified in the large-$N_c$ limit (large number of colors) to characterize nucleons properly 
\cite{Witten:1983tw,Adkins:1983ya,Zahed:1986qz}. 
When Skyrmions coexist with a domain wall (DW), the former are expected to be absorbed into the latter, forming a composite state called a DW-Skyrmion 
\cite{Nitta:2012wi,Nitta:2012rq,Gudnason:2014nba,Gudnason:2014hsa,Eto:2015uqa,Nitta:2022ahj}. 
Recently, they have been found to support a new phase, the DW-Skyrmion phase in QCD, in the presence of strong magnetic field or rapid rotation 
\cite{Eto:2023lyo,Eto:2023wul,Eto:2023tuu,Amari:2024mip,Amari:2024fbo,Amari:2025twm}.

Baby Skyrmions are two-dimensional analogues of Skyrmions supported by the second homotopy group 
$\pi_2 (S^2) \simeq {\mathbb Z}$ 
instead of the third one \cite{Piette:1994ug,Piette:1994mh}. 
Recently,  
magnetic Skyrmions have attracted great attention in condensed matter physics.  
Magnetic Skyrmions are 
particle-like solutions in the
magnetization vector of 
two-dimensional chiral magnetic 
materials 
\cite{1989JETP...68..101B,1995JETPL..62..247B,Nagaosa2013}. 
The magnetic Skyrmion owes its stability to the
Dzyaloshinskii-Moriya interaction (DMI) that is induced by the
spin-orbit coupling \cite{DZYALOSHINSKY1958241,Moriya:1960zz}, in contrast to the baby Skyrmion which is stabilized by the four derivative term, known as the Skyrme term. 
Magnetic Skyrmions are
important candidates for future storage systems in the 
form of race track memory, that potentially consumes much less energy
than current data storage technology \cite{Fert2013}, for a review see
\cite{Back:2019xvi}.
Magnetic Skyrmions were first realized experimentally in chiral
magnetic materials 
\cite{doi:10.1126/science.1166767,Yu2010}. 
Chiral magnets contain also other extended solutions that are
string-like, called magnetic DWs (domain lines) that separate magnetic
domains. 
In particular, in the plane there exists domain lines
\cite{doi:10.1126/science.1145799,KUMAR20221}.
Both the DW (or domain line in 2D) and the
magnetic Skyrmion are topologically protected objects, but the DW is stable and exists simply 
in the presence of 
the easy-axis anisotropy. 
Both the Skyrmion and the DW can be of two types: Bloch or
N\'eel, depending on how the magnetization vector twists around when
going from the center of the Skyrmion to its outside, whereas
the magnetization vector on a Bloch (N\'eel) DW is 
parallel (perpendicular) to the DW. 
In 3-dimensional
samples the magnetic domain wall is a surface object, and the magnetic Skyrmion extends as a (twisted) string-like
object through the material which can even form braids
\cite{Zheng2021}.
One of the difficulties in controlling Skyrmions under an applied current is the Skyrmion Hall effect \cite{Zang2011,wanjun2017,chen2017skyrmion}: the Skyrmion trajectory bends. 
To overcome this problem, 
it may be useful to consider Skyrmions 
trapped inside a DW where 
Skyrmions are restricted to move along the domain line. 
DW-Skyrmions in two dimensions were proposed in a field theoretical model 
\cite{Nitta:2012xq,Kobayashi:2013ju,Jennings:2013aea,Bychkov:2016cwc}. 
In the planar samples of chiral magnets, domain lines can stably host a Skyrmion \cite{Kim:2017lsi,PhysRevB.99.184412,PhysRevB.102.094402,Kuchkin:2020bkg,Ross:2022vsa,Lee:2022rxi,Amari:2023gqv,Lee:2024lge,Amari:2023bmx,Amari:2024jxx}, which creates a little cusp on the line. 
This object has been coined a magnetic DW-Skyrmion 
and has been discovered in the laboratory
\cite{Nagase:2020imn,10.1063/5.0056100,Yang2021}.
The domain-wall Skyrmions in chiral soliton lattices are unstable 
to decay into two merons \cite{Amari:2023bmx}, which are quasi-particles with topological charge $1/2$ \cite{Muller2017,Kharkov2017,Gobel2019,Mukai2022}.

In this paper, we find that capturing an isolated (stable) magnetic Skyrmion into a (stable) DW is not as easy as anticipated.
Indeed, we confirm with an exact analytic calculation of the asymptotic interactions as well as with detailed numerical computations, that the isolated Skyrmion in the bulk repels the DW.
In order to capture the isolated Skyrmion into the DW, we propose two ways of altering the solitons so as to make a successful capture possible: We change the complex phase of the DW or rotate the Skyrmion in the target space (2-dimensional isorotation).
By rotating the DW's phase into the unstable regime, we find that a successful capture can take place on separation distances of up to about 4.5 times the Skyrmion radius.
Rotating the Skyrmion, on the other hand, requires the Skyrmion to be in very close proximity to the DW, in order for the capture to be successful.
We also find that the stable Skyrmion becomes unstable when placed too close to a stable DW.
Finally, we discover a Kibble-like mechanism that can create both anti-DW-Skyrmions as well as multiple-DW-Skyrmion at the complex phase where the DW is maximally unstable. 

This paper is organized as follows.
In sec.~\ref{sec:model}, we introduce the magnetic Skyrmion model with the anisotropy potential.
In sec.~\ref{sec:setup}, we introduce the setup which is used as the initial condition for numerical computations and present the possible final states.
In sec.~\ref{sec:numerical}, we present the numerical computations that lead to the phase diagram from which one can read off where a successful capture of an isolated Skyrmion into a DW is possible.
In sec.~\ref{sec:interaction}, we compute the asymptotic interactions between the isolated Skyrmion and the DW.
Finally, we conclude in Sec.~\ref{sec:conclusion} with a discussion
and outlook.
We have relegated the connection between Bloch and N\'eel type DWs and
Skyrmions to appendix \ref{app:BlochNeel}.

\section{The chiral magnetic model}\label{sec:model}

The energy functional for a chiral magnetic material is given by
\beq
E=\int\calE\d^2x, \qquad
\calE=\frac12\p_i\bn\cdot\p_i\bn
+\kappa\bn\cdot\nabla\times\bn
+V(\bn),
\label{eq:E}
\eeq
with the anisotropy term
\beq
V(\bn)=\frac{1}{2}(1-n_3^2),
\label{eq:V}
\eeq
where we have rescaled length and energy units so as to only have dependence on one physical parameter, $\kappa$.\footnote{The physical constants of ref.~\cite{D0NR02947E} correspond to the value of the coupling $\kappa\approx 0.4$. }
The field $n$ is interpreted as the magnetization vector in chiral magnets, but here we take it as a nonlinear sigma model field with target space $S^2$: $\bn=(n_1,n_2,n_3)$ satisfying $\bn\cdot\bn=1$.

The magnetic Skyrmion carries a topological charge given by
\beq
Q = \int\calQ\;\d^2x,\qquad
\calQ = -\frac{1}{4\pi} \bn\cdot\p_1\bn\times\p_2\bn,
\label{eq:Q}
\eeq
which arises by considering the stereographic projection of the plane
to the 2-sphere, so that the magnetization vector is a map from the
2-sphere (the plane) to a unit vector in $\mathbb{R}^3$,
{\it i.e.}~another 2-sphere: this is characterized by
$\pi_2(S^2)=\mathbb{Z}\ni Q$, with $Q$ being the topological charge
given above.

The equation of motion corresponding to the energy \eqref{eq:E} is
\beq
\p_i^2n^a
- 2\kappa\epsilon^{aib}\p_in^b
+ n^3 \delta^{a3}
- \Lambda n^a = 0,
\eeq
where $\Lambda$ is a Lagrange multiplier used to impose the nonlinear sigma model constraint $\bn\cdot\bn=1$.
Changing to stereographic coordinates, $u$, the equation of motion becomes
\begin{align}
\mathrm{eom}&:=(1+|u|^2)\p_i^2u
-2\ub(\p_iu)^2
-2\kappa\left[\i(u+\ub)\p_1u +(u-\ub)\p_2u\right]
-u(1-|u|^2) = 0,
\label{eq:eomu}
\end{align}
where $u$ is defined as
\beq
u = \frac{n_1+\i n_2}{1+n_3},
\label{eq:u}
\eeq
which is defined everywhere on $S^2$, except at the south pole ($n_3=-1$).

\subsection{Setup and initial state}\label{sec:setup}

First, we will construct the composite initial configuration made of a
magnetic Skyrmion placed at the origin and a domain wall sitting at
$x=X_0<0$ with its world-line pointing in the $\hat{y}$-direction, see fig.~\ref{fig:setup}.
\begin{figure}[!htp]
\begin{center}
\begin{tikzpicture}[scale=0.2]
    \draw [<->,very thick] (17,18) -- (17,16) -- (19,16);
    \draw (15.5,17.5) node {$y$};
    \draw (19,14.5) node {$x$};
    \draw (-20,-20) rectangle (20,20);
    \filldraw [gray] (-10.5,-20) rectangle (-9.5,20);
    \draw (-10.5,-20) -- (-10.5,20);
    \draw (-9.5,-20) -- (-9.5,20);
    \filldraw [gray] (0,0) circle (0.92);
    \draw [<->,very thick] (-9.5,0) -- (-0.92,0);
    \draw (-5,2) node {\large $|X_0|$};
    \draw (2,-2.5) node {\large $(0,0)$};
\end{tikzpicture}
\caption{Setup of DW and isolated Skyrmion as initial condition \eqref{eq:u_composite}. }
\label{fig:setup}
\end{center}
\end{figure}

Each of the two solitons are solutions to the equations of motion if
they are well separated and the initial configuration is made using
the superposition rule in stereographic coordinates
\beq
u^{\rm composite} = u^{\rm sk} + u^{\rm DW},
\label{eq:u_composite}
\eeq
where the stereographic coordinate is defined in eq.~\eqref{eq:u}.
The magnetic Skyrmion and domain wall are given by
\begin{align}
  u^{\rm sk} = e^{\i(\theta+\beta)}\tan\left(\frac{f(r)}{2}\right),\qquad
  u^{\rm DW} = e^{\i\alpha-(x-X_0)},\label{eq:u_skyrmion_DW}
\end{align}
where the coordinate system in the plane is 
$x+\i y=r e^{\i\theta}$ with the Skyrmion placed at the origin and
the domain wall at $x=X_0$ 
parallel to the $\hat{y}$-direction.
We take $\kappa>0$ for which the stable magnetic Skyrmion and the stable DW correspond to
$\beta=\tfrac\pi2$ and $\alpha=\tfrac\pi2$, respectively.
The sign of the DW is chosen such that to the left of the DW ($x<X_0$)
the magnetization in the ground state is $n_3=-1$ while to the right of the
DW ($x>X_0$) $n_3=1$.
The well-separated solitons have energies
\begin{align}
  E^{\rm sk} &= \int\bigg[
    (f')^2
    +\sin^2f\left(1+\frac{1}{r^{2}}+4\kappa\sin\beta f'\right)
    \bigg]\pi r\d r,\non
  E^{\rm DW} &= \int\left[2 - \pi\kappa\sin\alpha\right]\d y,\label{eq:EDW}
\end{align}
from which it is clear that $\alpha=\frac\pi2$ is the minimal energy for
the DW, while for the Skyrmion $\beta=\frac\pi2$ is the minimal energy
too, since $f'=\frac{\d f}{\d r}$ is negative.
The DW solution ($u^{\rm DW}$) in eq.~\eqref{eq:u_skyrmion_DW} is exactly the renowned sine-Gordon solution in the inhomogeneous $\mathbb{C}P^1$ coordinate.

Inserting the Skyrmion field, $u^{\rm sk}$, of
eq.~\eqref{eq:u_skyrmion_DW} into the equation of motion \eqref{eq:eomu} yields
\begin{align}
f'' + \frac{f'}{r} - \frac{\sin2f}{2r^2}
+\frac{2\kappa\sin\beta\sin^2f}{r} - \frac{\sin2f}{2} = 0,
\label{eq:eomf}\\
\kappa\cos\beta\sin(f)f' = 0.
\end{align}
The first and the second equation correspond to the real and imaginary
part of the complex equation \eqref{eq:eomu}, respectively. 
The imaginary part fixes $\beta=\pm\frac\pi2$, with the upper sign
being the stable phase of the Skyrmion and the lower sign the unstable
phase.
Choosing the stable phase, $\sin\beta=1$ in eq.~\eqref{eq:eomf}.
Inserting instead the DW field, $u^{\rm DW}$, of eq.~\eqref{eq:u_skyrmion_DW} into the equation of motion \eqref{eq:eomu} yields
\beq
\kappa e^{-\i\alpha-2(x-X_0)}\cos\alpha = 0,
\eeq
which fixes the DW phase to be $\alpha=\pm\frac\pi2$, with the upper
sign being the stable phase of the DW and the lower sign the unstable
phase, as can also be seen from eq.~\eqref{eq:EDW}.

Note that the isolated Skyrmion is unstable for $\kappa\sin\beta<0$, which renders the DMI energy contribution positive.
This is a consequence of Derrick's theorem \cite{Derrick:1964ww}; Derrick's theorem states that solitons collapse in more than one spatial dimensions in a theory with a kinetic (gradient) term and a potential. 
The DMI term is a loophole, since it is not classically conformal invariant (like the kinetic term) and it is negative:
\beq
E^\lambda = E_2 + \lambda E_{\rm DMI} + \lambda^2 E_0,
\eeq
where $E_2$ is the kinetic (gradient) term, $E_{\rm DMI}$ is the DMI term, $E_0=\int V\d^2x$ is the potential term and $x\to\lambda x$ is a scale transformation.
For negative DMI energy, stabilization of the Skyrmion obeys the virial law ($\p E^\lambda/\p\lambda|_{\lambda=1}=0$):
\beq
E_{\rm DMI} = -2E_0,
\eeq
where $E_0>0$.
This is only possible when the DMI energy is negative and hence when $\beta=\pi/2$ in our notation.

We should emphasize that the superposition or ``Abrikosov Ansatz'' used in eq.~\eqref{eq:u_composite} is \emph{not} a solution, but simply an initial guess from which we will start numerical computations.
At asymptotic distances, however, the superposition becomes a very good approximation.
It is worth to point out that the superposition of two solitons in the inhomogeneous coordinates $u$ is very advantageous, since the composite soliton already satisfies the nonlinear constraint of the nonlinear sigma model via eq.~\eqref{eq:u}.

\subsection{Numerical results}\label{sec:numerical}

We will now turn to numerical computations.
In order to explore the possibility of a successful capture of an isolated Skyrmion into the DW, we perform
a large-scale set of simulations using the initial condition
\eqref{eq:u_composite} and evolving it to the nearest final state,
using the so-called arrested Newton flow algorithm.
The simulations are performed using a finite-distance approximation to
derivatives using a fourth-order 5-point stencil on a square lattice of
$682^2$ lattice sites.
The arrested Newton flow method is a second-order fictitious time-evolution (with fictitious time $\tau$) of the equations of motion (i.e.~akin to a wave equation)
\beq
\p_\tau^2u = \textrm{eom},
\eeq
with the special condition that the kinetic energy is removed (i.e.~$\p_\tau u=0$)
every time the potential part of the energy (i.e.~\eqref{eq:E}) increases and eom is given in eq.~\eqref{eq:eomu}.
The Newton flow accelerates towards the nearest local minimum of the energy functional \eqref{eq:E} and if not for the conditional arrest, it would blast right past it. Arresting it every time the flow has just passed the minimum ensures a fast convergence to the nearest local minimum. 
For simplicity, we use Neumann boundary conditions on the edges of the lattice, which
allows for the topological charges to enter or exit the simulation
area.

\subsubsection{The unsuccessful capture}

Naturally, as the first attempt we place the stable isolated Skyrmion (i.e.~with $\beta=\pi/2$) near the stable DW (i.e.~with $\alpha=\pi/2$); two outcomes can occur, either the Skyrmion is repelled (for large enough separation distance) or it is annihilated (for short enough separation distance).
This is bad news for the capture and creation of a DW-Skyrmion starting from an isolated Skyrmion and an empty DW.
We will get back to why the Skyrmion is annihilated by the DW shortly.

\subsubsection{Phase diagram and final states}

\begin{figure}[!htp]
\begin{center}
  \mbox{\includegraphics[height=0.2016\linewidth]{{\figsfolder}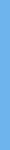}
  \includegraphics[width=0.2412\linewidth]{{\figsfolder}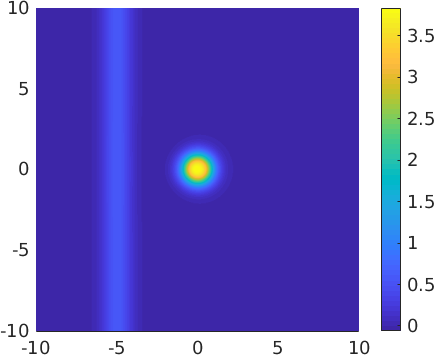}
  \includegraphics[width=0.2412\linewidth]{{\figsfolder}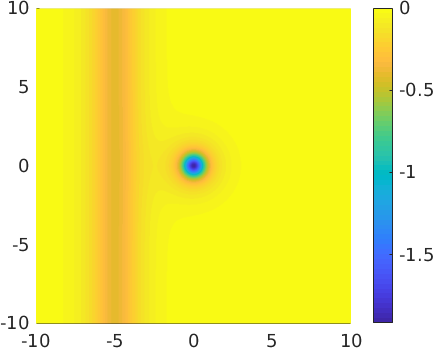}
    \includegraphics[width=0.23616\linewidth]{{\figsfolder}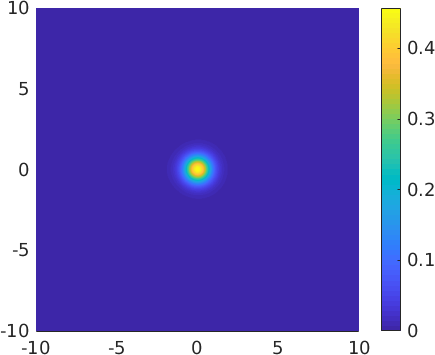}
    \includegraphics[width=0.19728\linewidth]{{\figsfolder}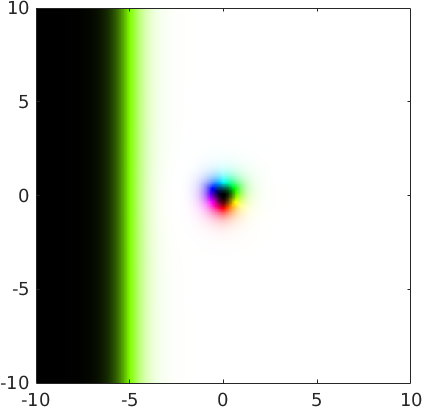}}
  \mbox{\includegraphics[height=0.2016\linewidth]{{\figsfolder}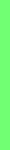}
  \includegraphics[width=0.2412\linewidth]{{\figsfolder}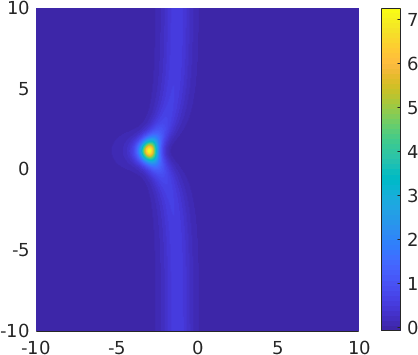}
  \includegraphics[width=0.2412\linewidth]{{\figsfolder}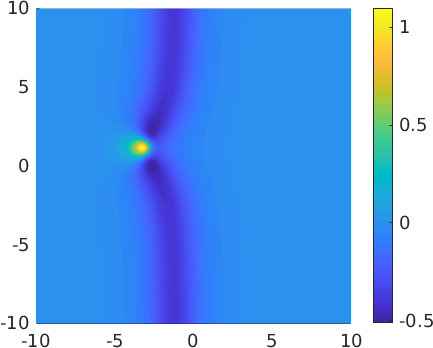}
    \includegraphics[width=0.23616\linewidth]{{\figsfolder}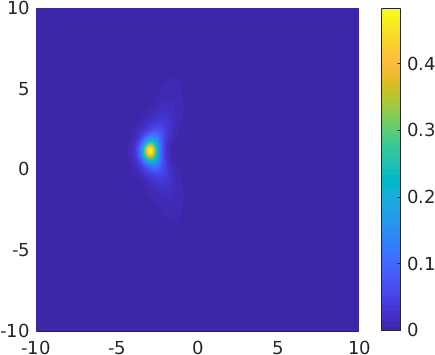}
    \includegraphics[width=0.19728\linewidth]{{\figsfolder}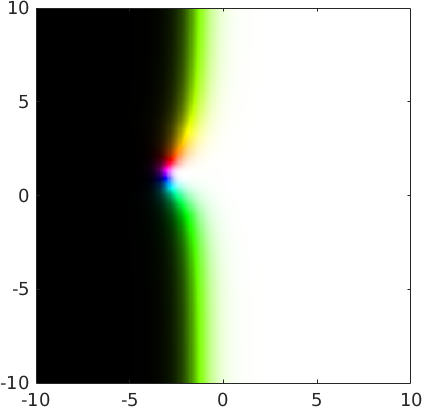}}
  \mbox{\includegraphics[height=0.2016\linewidth]{{\figsfolder}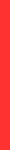}
  \includegraphics[width=0.23616\linewidth]{{\figsfolder}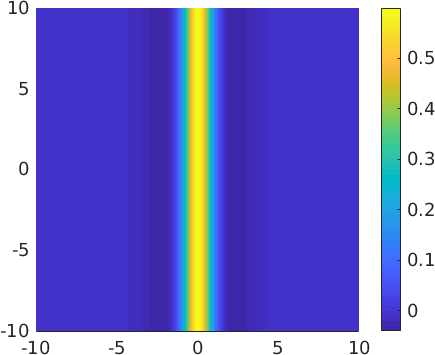}
  \includegraphics[width=0.2412\linewidth]{{\figsfolder}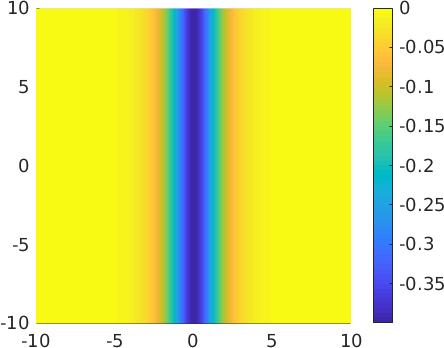}
    \includegraphics[width=0.2412\linewidth]{{\figsfolder}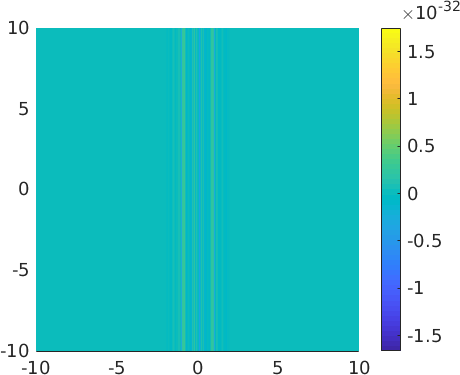}
    \includegraphics[width=0.19728\linewidth]{{\figsfolder}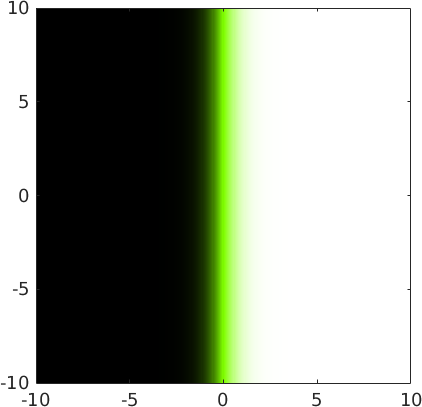}}
  \mbox{\includegraphics[height=0.2016\linewidth]{{\figsfolder}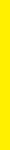}
  \includegraphics[width=0.2412\linewidth]{{\figsfolder}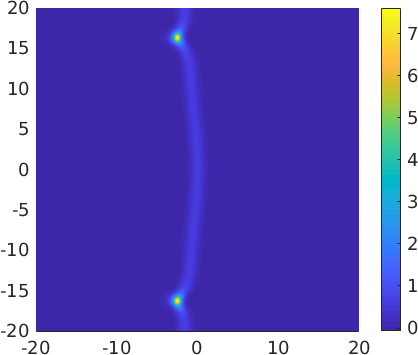}
  \includegraphics[width=0.2412\linewidth]{{\figsfolder}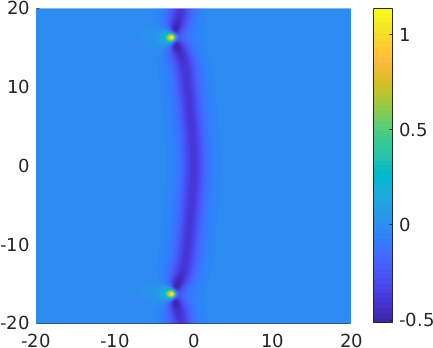}
    \includegraphics[width=0.23616\linewidth]{{\figsfolder}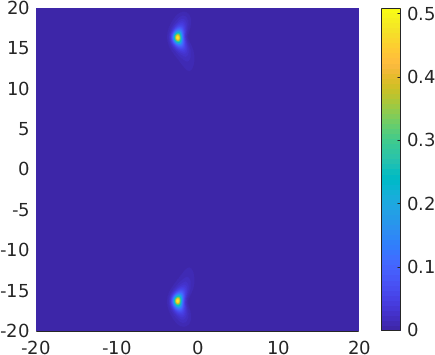}
    \includegraphics[width=0.19728\linewidth]{{\figsfolder}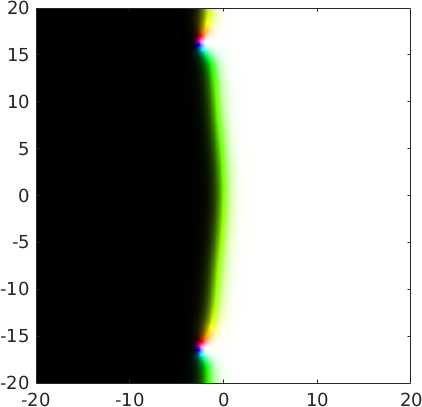}}
  \mbox{\includegraphics[height=0.2016\linewidth]{{\figsfolder}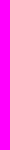}
  \includegraphics[width=0.2412\linewidth]{{\figsfolder}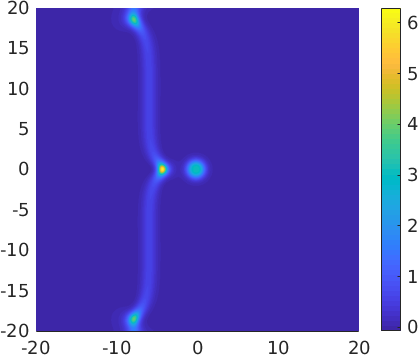}
  \includegraphics[width=0.2412\linewidth]{{\figsfolder}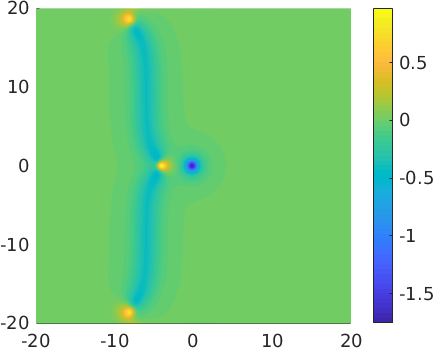}
    \includegraphics[width=0.23616\linewidth]{{\figsfolder}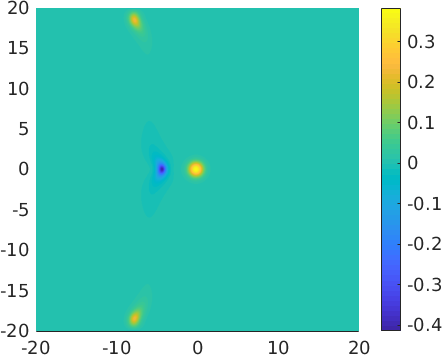}
    \includegraphics[width=0.19728\linewidth]{{\figsfolder}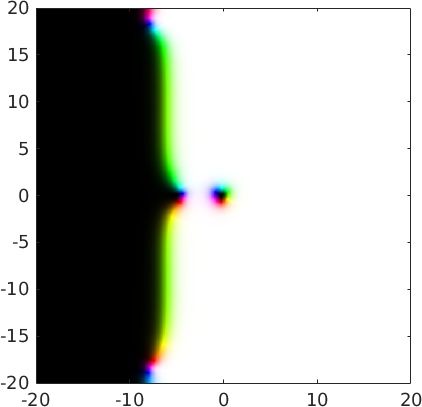}}
  \begin{tabular}{cccc}
    $\qquad\calE\qquad\qquad\qquad\qquad$&$\calE_{\rm DMI}\qquad\qquad\qquad$&$\calQ\qquad\qquad\qquad\qquad$&$\bn$
  \end{tabular}
  \end{center}
  \caption{The energy density ($\calE$), the DMI energy contribution to the energy ($\calE_{\rm DMI}$), the topological charge
    density ($\calQ$) and the magnetization vector $\bn$, for the
    five color-coded final states (colors shown left of each
    row): DW + Skyrmion (blue), DW-Skyrmion (green), only DW (red), anti-DW-Skyrmion and Skyrmion pair that annihilates and leaves behind two DW-Skyrmions (yellow) and finally the same as the previous case with the anti-DW-Skyrmion and bulk Skyrmion being a metastable bound state (magenta).
    These colors are referenced in fig.~\ref{fig:phasediagram}.
    The colors in the magnetization vector plot, $\bn$, (fourth column)
    are defined as: $n_3=1$ is white, $n_3=-1$ is black, $n_2=1$ is
    green, $n_1=1$ is red, $n_2=-1$ is magenta and $n_1=-1$ is cyan.
  }
  \label{fig:finalstates}
\end{figure}

\begin{figure}[!tp]
  \centering
  \mbox{\includegraphics[width=0.49\linewidth]{{\figsfolder}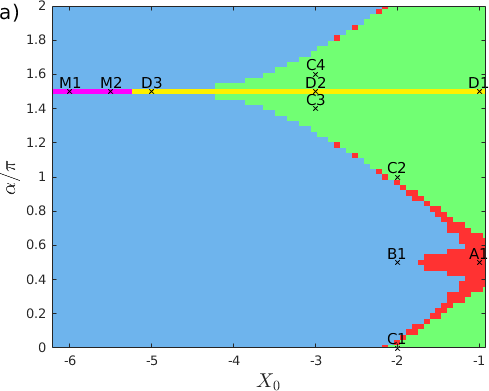}}
  \mbox{\includegraphics[width=0.49\linewidth]{{\figsfolder}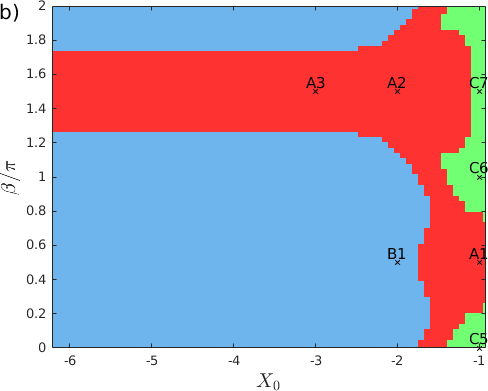}}
  \caption{The phase diagram of final states as a function of the
    initial parameters (a) $(X_0,\alpha)$ for $\beta=\frac\pi2$ and
    (b) $(X_0,\beta)$ for $\alpha=\frac\pi2$. The states are
    color coded with the same colors as in fig.~\ref{fig:finalstates}:
    DW + Skyrmion (blue), DW-Skyrmion (green), only DW (red), anti-DW-Skyrmion and Skyrmion pair that annihilates and leaves behind two DW-Skyrmions (yellow) and finally the same as the previous case with the anti-DW-Skyrmion and bulk Skyrmion being a metastable bound state (magenta).
  The magenta state eventually decays into the yellow state and the yellow state can decay into the red state for finite specimen sizes.
  The marked points with (e.g.~A1,B1,$\cdots$) correspond to videos available in the supplemental material.
  }
  \label{fig:phasediagram}
\end{figure}

In order to search for a successful capture, we perform a scan over the phase parameters $\alpha$ and $\beta$ for a range of separation distances $X_0$.
The phase diagram computed by running one
simulation for each point in a lattice of initial condition parameters
with 64 points parametrizing the angle $\alpha$ ($\beta$) and $74$ points
parametrizing the separation distance in the range $X_0\in[-6.2,-R]$,
with $R\approx 0.92$ being the Skyrmion radius.
This yields a grand total of $4736$ simulation for the $\alpha$-phase
diagram with $\beta=\frac\pi2$ and the same number for varying $\beta$
at fixed $\alpha=\frac\pi2$.
Due to the large number of simulations, each with a large number of
flops, we used our CUDA C code on a GPU cluster.
The final states are shown in fig.~\ref{fig:finalstates} and the resulting phase diagram, displaying which final state is reached by the numerical computation for the given values of $\alpha$, $\beta=\frac{\pi}{2}$ and $X_0$ ($\alpha=\frac{\pi}{2}$, $\beta$ and $X_0$) are shown in fig.~\ref{fig:phasediagram}a (fig.~\ref{fig:phasediagram}b).

The initial condition $u^{\rm composite}$ is in general unstable and
each numerical computation will flow the specific initial configuration with $(\alpha,\beta,X_0)$ to its nearest local minimum of the energy functional, which we shall denote final states.
This results in 5 different final states, which we show in fig.~\ref{fig:finalstates} (of which 2 are metastable): 
\begin{itemize}
\item the separated magnetic Skyrmion and DW with increasing separation 
distance (they repel each other in their respective stable stable states (blue color code),
\item the DW-Skyrmion where the magnetic Skyrmion has been captured into the world 
volume of the DW (green color code),
\item the annihilation of the magnetic Skyrmion by shrinkage (red color code),
\item a metastable state with an anti-Skyrmion induced on the DW by the long-distance forces from a Skyrmion in the bulk, which quickly annihilates and leaves a pair of DW-Skyrmions on the DW (yellow color code),
\item same as the above-described state, but with the anti-DW-Skyrmion and the bulk Skyrmion being a long-lived composite before the eventual annihilation (magenta color code).
\end{itemize}

Starting with the stable Skyrmion ($\beta=\frac\pi2$), 
we obtain, as promised, that the stable DW ($\alpha=\frac\pi2$) annihilates the Skyrmion for separation distances smaller than $|X_0|\lesssim1.75$ whereas the Skyrmion is repelled by the DW for larger separation distances.
In order to capture the isolated stable Skyrmion into the DW for the creation of the DW-Skyrmion composite soliton, one approach that works and is found for the first time in this paper, is by modifying the phase of the DW so that it is not in its stable phase (ground state). 
This corresponds to either increasing or decreasing $\alpha$ from $\alpha=\frac\pi2$.
The unstable DW phase is not enough by itself, as the DW will simply flow back to its stable position (ground state); hence, it is also necessary that the Skyrmion is in close proximity to the DW at the time when the DW phase is altered from its stable value. 
This capture corresponds to the green-colored area in fig.~\ref{fig:phasediagram}a, which forms a wedge in both directions away from the stable $\alpha=\frac\pi2$ phase.
The longest distance for which the Skyrmion can be captured by the DW is found near the most unstable phase of the DW, i.e.~$\alpha=\frac{3\pi}{2}$ (with the exception for the exact antipodal value $\alpha=\frac{3\pi}{2}$ where an array of other phenomena appears, see below).

Turning now to the most unstable point of the DW, i.e.~$\alpha=\frac{3\pi}{2}$, the DW is now on the top of the hill in the static energy landscape and has the potential to create Skyrmion-anti-Skyrmion pairs. 
This is simply the 1-dimensional version of the Kibble mechanism \cite{Kibble:1976sj,Kibble:1980mv}. 
Now, instead of random perturbations, as in the cosmological setting of the Kibble mechanism, here the perturbation is specific and is given by the interaction between the DW and the (isolated) Skyrmion.
For separation distances shorter than $|X_0|\lesssim5$, the perturbation of the unstable DW creates an anti-Skyrmion on the DW, but the anti-DW-Skyrmion is in such close proximity to the isolated Skyrmion in the bulk, that the two solitons annihilate and disappear.
In the process of perturbing the DW, the perturbation also creates a pair of DW-Skyrmions on each side of the anti-DW-Skyrmion.
The minimization of energy, however, pushes them away and off of our finite-sized computation grid.
Increasing the separation distance ($|X_0|$), at the unstable DW phase (i.e.~$\alpha=\frac{3\pi}{2}$), still creates an anti-DW-Skyrmion by perturbing the unstable DW, but now the bulk (isolated) Skyrmion is so far away that they remain as a metastable bound state for some time -- until they eventually come close enough to each other to annihilate.

We now turn to the more physically difficult case of leaving the DW in its stable phase and rotating the isolated Skyrmion (in target space coordinates).
In this case, the isolated Skyrmion becomes unstable when the phase is $\beta=\frac{3\pi}{2}$ for which the DMI energy is positive and hence there is no loophole to avoiding Derrick's theorem.
When the (isolated) Skyrmion is sufficiently far from the DW, indeed this is what we see: the Skyrmion collapses and disappears (see the red stripe at large $|X_0|$ at $\beta=\frac{3\pi}{2}$ in fig.~\ref{fig:phasediagram}b).
We already know that for a stable (isolated) Skyrmion the (stable) DW repels or annihilates the Skyrmion, so the only possibility of capture is when the Skyrmion is rotated by some angle away from $\beta=\frac\pi2$.
Indeed, this is what we see from fig.~\ref{fig:phasediagram}b.
We find, however, that only when the (isolated) Skyrmion is in very close proximity to the DW, a successful capture is possible, see the green-shaded area in fig.~\ref{fig:phasediagram}b.

It is clear that when the isolated Skyrmion in the bulk is rotated into the unstable phase ($\beta=\frac{3\pi}{2}$), it decays in the absence of interaction with the DW, see the large-$|X_0|$ (left) part of fig.~\ref{fig:phasediagram}b.

\begin{figure}[!htp]
\centering
\mbox{\raisebox{40pt}{$\ \ \bn\ \ $}\includegraphics[width=0.195\linewidth]{{\figsfolder}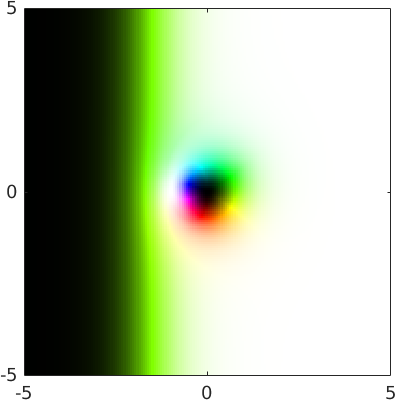}
\includegraphics[width=0.195\linewidth]{{\figsfolder}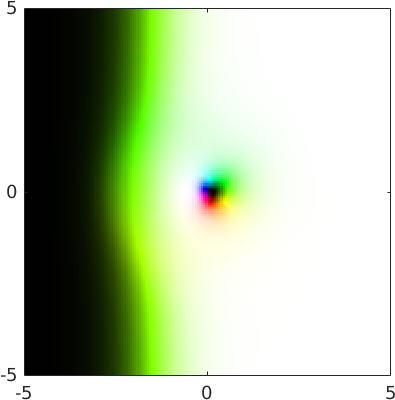}
\includegraphics[width=0.195\linewidth]{{\figsfolder}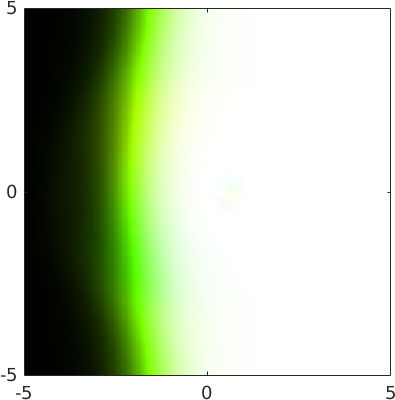}}
\mbox{\raisebox{40pt}{$\calE_{\rm DMI}$}\includegraphics[width=0.195\linewidth]{{\figsfolder}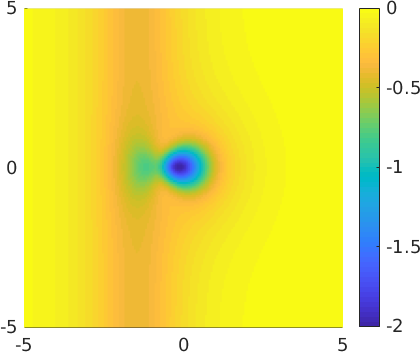}
\includegraphics[width=0.195\linewidth]{{\figsfolder}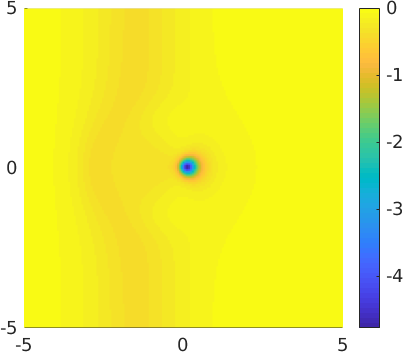}
\includegraphics[width=0.195\linewidth]{{\figsfolder}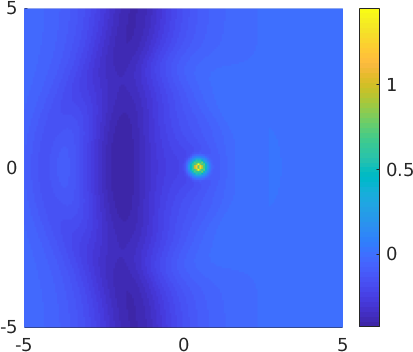}}
\caption{Shrinkage of a Skyrmion in the near proximity to the DW.
The left column corresponds to the initial configuration \eqref{eq:u_composite} (with $X_0=-1.5$), the middle to arrested Newton flow (aNF) step number 150 and the right-most column to step number 250.
In this figure $\alpha=\beta=\frac\pi2$ and $\kappa=0.4$.}
\label{fig:shrink}
\end{figure}
The collapse of the stable (isolated) Skyrmion (i.e.~with $\beta=\frac\pi2$) at close proximity to the DW can be explained as follows. 
The DMI energy of the isolated Skyrmion is negative (see the second column of the first row (blue) in fig.~\ref{fig:finalstates}), whereas the DMI energy of the DW-Skyrmion is positive (see the second column of the second row (green) in fig.~\ref{fig:finalstates}).
When there is enough ``momentum'' to overcome the barrier from going from an isolated bulk Skyrmion to becoming a DW-Skyrmion, it works out. 
If there the Skyrmion gets ``stuck'' midway, it suffers the issue of being repelled and gaining a negative DMI energy or being attracted and gaining a positive DMI energy. 
With a nearly vanishing DMI energy, there is no loophole to Derrick's theorem and the soliton collapses.
Hence the demise of the Skyrmion, see fig.~\ref{fig:shrink}.

The final issue that needs more theoretical explanation, is why does a stable (isolated) Skyrmion and a stable DW repel each other?
We will explain this using an asymptotic analytic calculation in the next subsection.

\subsection{Interactions}\label{sec:interaction}

The aim of this section is to understand, qualitatively, why the stable DW [$\alpha=\frac\pi2$] and the stable (isolated) Skyrmion [$\beta=\frac\pi2$] repel each other at large separation distances.
This can be understood by studying the asymptotic interactions.

The analytic computation of the interaction energy is found as
follows.
We separate the plane into two regions: Left and right and place the
DW on the left and the magnetic Skyrmion on the right.
Each field is assumed to be small in the other soliton's core region
and we assume the soliton in its own core region to obey the equation
of motion.
The interaction energy is defined as the total energy with each of the
individual solitons' energies subtracted off 
\beq
E^{\rm int}=E^{\rm DW+sk}-E^{\rm DW}-E^{\rm sk}.
\label{eq:Eint}
\eeq
Expanding the energy \eqref{eq:E} around a
solution as $u\to u+\du$, the expression for the
interaction energy will simplify as the part of the energy of the soliton (DW or
Skyrmion) will cancel out.
In order to perform the computation, however, it is important that the
variational problem is well defined, which as described in
refs.~\cite{Doring2017,Barton-Singer:2018dlh,Bolognesi:2024mjs}
requires that we drop a boundary term in the energy functional.
The equation of motion is of course unchanged by dropping this
boundary term.
Integrating the energy \eqref{eq:E} by parts, we identify the boundary
term that should be dropped
\begin{align}
\calE &= \frac12\p_i\bn\cdot\p_i\bn
+\kappa\epsilon^{aib}n^a\p_in^b
+\frac12[1-(n^3)^2]\non
&= \frac12\p_i\bn\cdot\p_i\bn +
2\kappa\epsilon^{ij}n^i\p_jn^3
+\frac12\left[1-(n^3)^2\right] +
\kappa\epsilon^{ij}\p_i(n^j n^3), \qquad i,j=1,2.
\end{align}
Expanding the energy functional without the boundary term (i.e.~the
last term) about the perturbations, we obtain
\begin{align}
\calE &= \calE_0
- \frac{2\,\overline{\mathrm{eom}}}{(1+|u|^2)^3}\du
+\p_i\left(\left[\frac{2\p_i\ub}{(1+|u|^2)^2} - \frac{4\i\kappa(\vb_i u-v_i\ub)\ub}{(1+|u|^2)^3}\right]\du\right)
+\textrm{c.c.}
+\calO(\du^2,\dub^2),
\end{align}
where $v = (1,\i)$.
The equation of motion (eom) vanishes on the background solution.
Hence, to linear order we obtain\footnote{Changing the phase $\alpha$
for the DW or the phase $\beta$ for the Skyrmion away from the stable
point $(\alpha=\beta=\frac\pi2)$ corresponds to turning on a quadratic
field operator in the piece eom, which we neglect to leading order. }
\begin{equation}
\calE = \calE_0
+2\p_i(\p_i\ub\du)
+2\p_i(\p_iu\dub)
+\calO(\du^2,\dub^2).
\end{equation}

Specializing now to the case of the DW sitting on the left and the
Skyrmion on the right, we may write the interaction energy as
\begin{align}
  E^{\rm int} &= 2\int_L\p_i\left[\p_iu^L\ub^R+\textrm{c.c.}\right]\d^2x
  +2\int_R\p_i\left[\p_iu^R\ub^L +\textrm{c.c.}\right]\d^2x\non
  &=2\int_\Gamma\left[
    \p_iu^L\ub^R
    -\p_iu^R\ub^L
    +\textrm{c.c.}
    \right]\d S^i,
  \label{eq:Eint1}
\end{align}
where $\Gamma$ is a line in the $\hat{y}$-direction, separating the DW
and the Skyrmion.
In the left region, we use $u=u^L=u^{\rm DW}$ as the background
solution with $\du=u^R=u^{\rm sk}$ being the perturbation
and in the right region, we use $u=u^R=u^{\rm sk}$ as the
background solution with $\du=u^L=u^{\rm DW}$ being the
perturbation.
We assume that both $u^L=u^{\rm DW}$ and $u^R=u^{\rm sk}$ are
small on the line $\Gamma$ that separates the left and the right
regions.

In order to evaluate the line integral in eq.~\eqref{eq:Eint1},
we use Green's theorem and turn the integral into an area integral
over the right-hand domain, that includes the Skyrmion
\begin{align}
  E^{\rm int} &=-2\int_R\left[
    \Delta u^L\ub^R
    -\Delta u^R\ub^L
    +\textrm{c.c.}
    \right]\d^2x.
  \label{eq:Eint2}
\end{align}
Since this is a mathematical trick to evaluate the line integral on
$\Gamma$, we can now use that both $u^L$ and $u^R$ can be described by
their exact solutions to their respective linearized equations of
motion.
Since $u^L=u^{\rm DW}$ is already exact and exponentially small in the
right region, this applies to the Skyrmion field $u^R=u^{\rm sk}$.
Linearizing the equation of motion for the Skyrmion \eqref{eq:eomf},
we get
\beq
f'' + \frac{f'}{r} - \frac{f}{r^2} - f = 0,
\eeq
which has the exact solution $f = q K_1(r)$,
and in turn
\beq
u^{\rm sk,\,linear} \simeq \frac{q}{2}e^{\i(\theta+\beta)}K_1(r)
=-q e^{\i\beta}\p_{\zb}K_0(r),
\eeq
where $K_1(r)$ and $K_0(r)$ are modified Bessel functions of the
second kind.
Comparing with the 2D Green function
$(\Delta - 1)K_0(r) = -2\pi\delta^{(2)}(\bx)$,
we get that the equation for the asymptotic solution
$u^{\rm sk}$ is
\beq
(\Delta-1)u^R = 2\pi q e^{\i\beta}\p_{\zb}\delta^{(2)}(\bx),
\label{eq:aeomuR}
\eeq
which has a delta-function source at its origin \cite{Piette:1994ug,Foster:2019rbd}.
This linearized solution is valid on the line $\Gamma$ and the line
integral is by Green's theorem equal to the area integral using
this delta-function sourced linearized solution.
The DW field, on the other hand, is not sourced in the right region,
hence
\beq
(\Delta-1)u^L = 0.
\label{eq:aeomuL}
\eeq
Using eqs.~~\eqref{eq:aeomuR} and \eqref{eq:aeomuL}, we obtain
\begin{align}
  E^{\rm int} &=2\int_R\left[
    \ub^L(\Delta -1) u^R
    +\textrm{c.c.}
    \right]\d^2x\non
  &=4\pi q \cos(\beta-\alpha) e^{-|X_0|},
  \label{eq:Eint3}
\end{align}
with $q$ being a positive constant, $\alpha$ and $\beta$ are the
phases of the DW and the Skyrmion, respectively, and finally $X_0(<0)$
is the DW position.
We see from the analytic leading-order result, that the stable
Skyrmion and stable DW ($\beta=\alpha=\frac\pi2$) yield a positive
interaction energy, corresponding to repulsion.
On the other hand, if either the DW or the Skyrmion is rotated away
from their respective stable phases, attraction is possible.
This explains why the well separated stable DW and stable (isolated) Skyrmion repel each other.
At short distances, this asymptotic interaction cannot be trusted as nonlinear terms become important for the minute details of the interactions.

\section{Conclusion and outlook}\label{sec:conclusion}

In this paper, we have investigated the capture of a single isolated Skyrmion into a stable DW in a chiral magnet.
It turns out that they repel each other at large distances and at short distance the Skyrmion is obliterated by the following effect: The stable isolated Skyrmion in the bulk has a negative DMI energy, whereas the stable DW-Skyrmion has a positive DMI energy. At short distances, if the Skyrmion does not have a momentum to overcome a certain (nonlinear) barrier, the DMI energy remains small or positive leading to a shrinking instability of the Skyrmion, unless the Skyrmion manages to be pushed all the way into the DW and become a stable DW-Skyrmion -- a cuspy kink of twisted spins as a sine-Gordon soliton living on the DW.
In this paper we propose two ways to obtain a successful capture of the isolated Skyrmion in the bulk into the DW: Either we perturb the DW into an unstable complex phase (away from its minimum of the energy) or we rotate the Skyrmion on the target space, which increases its DMI energy (from a negative value towards zero). 
Both perturbations can lead to a successful capture of the isolated Skyrmion into the DW, provided that the Skyrmion is in close enough proximity to the DW. 
This is mapped out in the phase diagram shown in fig.~\ref{fig:phasediagram}.
When the DW is perturbed to the maximally unstable complex phase ($\alpha=\frac{3\pi}{2}$), it can fall into the stable phase ($\alpha=\frac\pi2$) in two different directions. 
This is akin to the 1-dimensional version of the Kibble mechanism, which indeed can create DW-Skyrmions or anti-DW-Skyrmions.
As opposed to the Kibble mechanism, the perturbations are not random, but are indeed given by the asymptotic interaction between the DW and the Skyrmion, which creates an anti-DW-Skyrmion on the DW facing the bulk Skyrmion.
This pair of solitons annihilate and disappear, except for a large enough separation distance, where they appear as a metastable bound state for quite some time, before they eventually collide and annihilate.
Finally, on the theoretical side we show analytically that the asymptotic interaction between an isolated Skyrmion and a DW leads to a repulsive force.

With our proposals, we have found two ways to capture an isolated Skyrmion into a DW, i.e.~by rotating the DW phase or by rotating the Skyrmion on the target space, but there could be many more ways to perform a successful capture -- some of all these could become experimentally viable in real chiral magnets.
Although we performed the analysis using the Bloch DMI by convention, we could equivalently have used a N\'eel DMI instead (see Appendix \ref{app:BlochNeel}).

A future direction of study would be to use the Landau-Lifshitz-Gilbert (LLG) equation instead of energy minimization (here arrested Newton flow). 
An LLG flow without a current turned on corresponds to a gradient flow, which is essentially an exponentially slower version of the arrested Newton flow, but with an added symplectic flow that pushes the ``particles'' in the flow in the orthogonal direction to the flow. 
For a field and hence a soliton, this has consequences modifying the phase diagram of the final states.
We expect shifts in the phase diagram in the angular directions (i.e.~$\alpha$ or $\beta$), but otherwise qualitatively similar behaviors up to angular shifts.
Another interesting direction for future studies would be to find a way to start with a DW-Skyrmion and somehow eject the Skyrmion so that one ends up with an empty DW and an isolated Skyrmion -- viz.~the inverse process of what we have studied in this paper.

\subsection*{Note added}

After posting this paper on the arXiv, a preprint with some overlap
also appeared on the arXiv \cite{Leask:2024dlo}. 
In Ref.~\cite{Leask:2024dlo}, configurations similar to those we display in Fig.~\ref{fig:finalstates} are found and a configuration with two DWs is also studied.
Instead, in our paper we provide a systematic result in form of the phase diagram shown in Fig.~\ref{fig:phasediagram} and we compute the asymptotic interactions between the DW and the isolated magnetic Skyrmion analytically.

\subsection*{Acknowledgments}
S.~B.~G.~thanks the Outstanding Talent Program of Henan University for partial support.
This work is supported in part by JSPS KAKENHI [Grants No.~JP23KJ1881
  (Y.~A.), No.~JP22H01221 and  No.~JP23K22492 
  (M.~N.)] and the WPI program
``Sustainability with Knotted Chiral Meta Matter (WPI-SKCM$^2$)'' at
Hiroshima University (M.~N.).

\appendix
\numberwithin{equation}{section}

\section{Bloch versus N\'eel DW-Skyrmion}\label{app:BlochNeel}

We can formally consider the static energy of the form of a gauged
nonlinear sigma model
\cite{li2011general,Schroers:2019hhe,Amari:2023gqv}
\begin{align}
  \calE &= \frac12D_i\bn\cdot D_i\bn + m^2\big(1-(n^3)^2\big)\non
  &= \frac12\p_i\bn\cdot\p_i\bn + \frac12\bA_i\cdot\bA_i -
  \frac12(\bA_i\cdot\bn)^2 + \p_i\bn\cdot\bA_i\times\bn + m^2\big(1-(n^3)^2\big),
\label{eq:E_GNLSM}
\end{align}
with the gauge covariant derivative
\beq
D_i\bn = \p_i\bn + \bA_i\times\bn,
\eeq
and $\bn=(n^1,n^2,n^3)$ is the magnetization field as usual.
We choose a constant background gauge field ($\bA_i=(A_i^1,A_i^2,A_i^3)$) as
\beq
A_i^a =
\begin{cases}
  -\kappa\delta_i^a,\qquad & \textrm{Dresselhaus SOC},\\
  -\kappa\epsilon_i^{\phantom{i}a}, & \textrm{Rashba SOC}.
\end{cases}
\eeq
In both cases, the field strength is
\beq
F_{12} = F_{12}^a\frac{\tau^a}{2} = \kappa^2\tau^3,
\eeq
implying that the above two choices are transformed to each other by a gauge transformation, given below.
Inserting the two choices of gauge field into the energy functional
\eqref{eq:E_GNLSM} yields
\begin{align}
\calE &=
\frac12\p_i\bn\cdot\p_i\bn + \kappa^2
+ \left(m^2-\frac{\kappa^2}{2}\right)\big(1-(n^3)^2\big)\non&\phantom{=\ }
+
\begin{cases}
  \kappa\bn\cdot\nabla\times\bn, & \textrm{Dresselhaus SOC},\\
  \kappa\left(\bn\cdot\nabla n^3 - n^3\nabla\cdot\bn\right),\qquad & \textrm{Rashba SOC}.
\end{cases}
\end{align}

For the gauge field corresponding to the Dresselhaus SOC, the
Lagrangian is simply given by eq.~\eqref{eq:E_GNLSM} up to a constant ($\kappa^2$) with $2m^2-\kappa^2=1$.
The other gauge field gives rise to a seemingly different form of the
DM term.
It is known that the Dresselhaus-SOC-type DM term gives rise to the
Bloch-type DW-Skyrmion, whereas the Rashba-SOC-type DM term gives rise
to the N\'eel-type  \cite{Amari:2023gqv}.
On the other hand, in the energy functional \eqref{eq:E_GNLSM}, the
difference between the two types of DM term is simply a gauge choice.
In fact, a gauge transformation
\begin{align}
  A_i^a &\to R^{ab} A_i^b + \i\tr[\tau^a U^{-1}\p_i U],\\
  n^a &\to R^{ab} n^b, \qquad
  R^{ab} = \frac12\tr[\tau^a U\tau^b U^{-1}],
\end{align}
with $U\in\SU(2)$ and $R\in\SO(3)$ leaves the energy functional
\eqref{eq:E_GNLSM} invariant.
The following gauge transformation
\beq
U = 
\begin{pmatrix}
  e^{-\frac{\i\pi}{4}} & 0\\
  0 & e^{\frac{\i\pi}{4}}
\end{pmatrix}, \qquad
R^{ab} =
\begin{pmatrix}
  0 & -1 & 0\\
  1 & 0 & 0\\
  0 & 0 & 1
\end{pmatrix},
\eeq
changes the gauge field from the Dresselhaus SOC type to the Rashba
SOC type and in turn $\bn=(n^1,n^2,n^3)\to(-n^2,n^1,n^3)$.
This transformation maps a Bloch DW to a N\'eel DW.

\bibliographystyle{JHEP}
\bibliography{references}

\end{document}